# Phonon transport in single-layer $Mo_{1-x}W_xS_2$ alloy embedded with $WS_2$ nanodomains

Xiaokun Gu[1] and Ronggui Yang[1,2,*]

[1]Department of Mechanical Engineering, University of Colorado, Boulder, Colorado, 80309, USA

[2]Materials Science and Engineering Program, University of Colorado, Boulder, Colorado, 80309, USA

*ronggui.yang@colorado.edu

**Abstract**

Two-dimensional (2-D) transition metal dichalcogenides (TMDs) have shown numerous interesting physical and chemical properties, making them promising materials for electronic, optoelectronic, and energy applications. Tuning thermal conductivity of two-dimensional (2-D) materials could expand their applicability in many of these fields. In this paper, we propose a strategy of using alloying and nanodomains to suppress the thermal conductivity of 2-D materials. To predict the thermal conductivity of 2-D alloy embedded with nanodomains, we employ the Green's function approach to assess the phonon scattering strength due to alloying and nanodomain embedding. Our first-principles-driven phonon Boltzmann transport equation calculations show that the thermal conductivity of single-layer $MoS_2$ can be reduced to less than 1/10 of its intrinsic thermal conductivity after alloying with W and introducing nanodomains due to the strong scattering for both high- and low-frequency phonons. The strategies to further reduce the thermal conductivity are also discussed.

## I. Introduction

Two-dimensional (2-D) transition metal dichalcogenides (TMDs) have shown numerous interesting physical and chemical properties,[1-3] making them promising materials for electronic, optoelectronic, and energy applications. For many of these technological applications, materials are expected to possess a few desired functions or properties simultaneously, which can hardly be fulfilled by the intrinsic properties of a single material. Therefore, various attempts, such as reducing dimensions,[4,5] intercalation,[6,7] heterostructuring[8-10] and alloying,[11,12] have been made to tune the electronic and optical properties of 2-D TMDs to expand the applicability of 2-D TMDs for different applications.

Phonon and thermal properties of 2-D TMDs is of great interest because it is highly relevant to the functionality, performance and reliability of 2-D TMD-enabled devices. Many efforts have been devoted to investigate the thermal conductivity of 2-D TMDs, both theoretically[13-15] and experimentally[16-19]. While the 2-D TMDs with high thermal conductivity might be beneficial to thermal management and electronic cooling, those with low thermal conductivity could be used as the thermal barrier materials[20] and thermoelectric materials.[21] Tuning the thermal conductivity of 2-D TMDs could significantly broaden their applications.[22] For example, recent experiments showed that $MoS_2$ is of a relatively high power factor.[23,24] Apparently if the thermal conductivity of 2-D $MoS_2$ can be suppressed without significant change in power factor, it could be a promising thermoelectric material.

Inspired by the so-called "nanoparticle-in-alloy" approach used to reduce the thermal conductivity of three-dimensional (3-D) bulk materials, *i.e.,* nanocomposites,[25-27] one might expect very low thermal conductivity of 2-D TMD alloys when embedded with nanodomains. Interestingly, both 2-D $MoS_2$-based alloys[11,12] and heterostructures with triangular nanodomains have been recently synthesized,[8-10] which laid the foundation for synthesizing 2-D TMD alloys embedded with nanodomains. However, it is unclear how low the thermal conductivity can be achieved using the "nanodomains in 2-D alloy" approach. This calls for a fundamental study on how alloying and embedding nanodomains affect phonon transport and thermal conductivity of 2-D TMDs.

In this paper, we study the lattice (phonon) thermal conductivity of single-layer $Mo_{1-x}W_xS_2$ alloy and $Mo_{1-x}W_xS_2$ alloy embedded with triangular $WS_2$ nanodomains. Since $MoS_2$ and $WS_2$ are almost lattice matched,[13] the nanostructures we studied here could be dislocation free and are expected to retain a relatively high power factor without significantly shortening electron mean free paths.[25-27] The first-principles-based Pierels-Boltzmann transport equation (PBTE) approach is employed to calculate the thermal conductivity of the nanostructures. The phonon scattering mechanisms, including three-phonon scattering, phonon-alloy scattering and phonon-nanodomain scattering, are all accounted for in the solution of the PBTE. The phonon scattering rates due to both alloying and embedding nanodomains are evaluated by the Green's function approach.[28,29] The effects of area fraction, nanodomain size and the composition of alloy on the thermal conductivity are investigated. The thermal conductivity was found to be reduced to one-tenth of the defect-free single-layer $MoS_2$ mainly due to the strong phonon-alloy scattering. Nanodomains can also play a role in thermal conductivity reduction.

## II. Numerical Method

### A. Pierels-Boltzmann transport equation for lattice thermal conductivity

Figure 1 shows a single-layer $Mo_{1-x}W_xS_2$ alloy embedded with triangular $WS_2$ nanodomains lying in the *x-y* plane. A typical atomic configuration for the alloy with nanodomains is also presented in Fig. 1. Assuming that all the nanodomains are of the same size but randomly embedded, the 2-D crystal is then characterized by three parameters, including the composition of alloy, $x$, the area fraction, $f^{\mathrm{ND}}$, and the side length, $a$, of $WS_2$ nanodomains. $f^{\mathrm{ND}}/A^{\mathrm{ND}}$ defines the number density of nanodomains per unit area $N^{\mathrm{ND}}$, where $A^{\mathrm{ND}}$ is the area of each nanodomain.

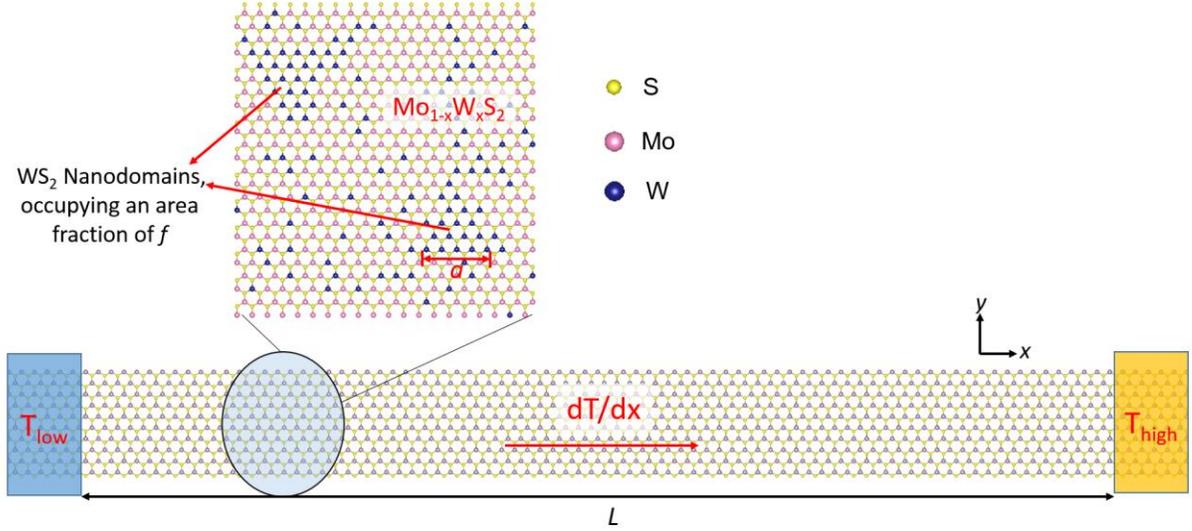

*Figure 1. The schematic of the simulation domain. The inset shows a typical atomic configuration for the $Mo_{1-x}W_xS_2$ alloy embedded with triangular $WS_2$ nanodomains.*

To calculate the in-plane thermal conductivity of the 2-D materials, a small temperature difference $\Delta T$ is applied to the two ends of the monolayer with a distance $L$ apart, resulting in a temperature gradient along $x$ direction, $dT/dx$. The effective thermal conductivity can then be expressed as the summation of the thermal conductivity contributed by each phonon mode in first Brillouin zone, and is written as[30,31]

$$\kappa^{xx} = \frac{1}{(2\pi)^2 h} \sum_s \int \hbar \omega_{\mathbf{q}s} v_{\mathbf{q}s}^x n_{\mathbf{q}s}^0 (n_{\mathbf{q}s}^0 + 1) F_{\mathbf{q}s} d\mathbf{q}, \tag{1}$$

where $h$ is the thickness of the 2-D crystal, $\hbar$ is the Planck constant, $\mathbf{q}s$ stands for the $s$-th phonon mode with $\hbar\mathbf{q}$ momentum, $\omega$, $v$ and $n^0$ are the phonon frequency, group velocity and equilibrium phonon population, respectively, which are determined by the phonon dispersion relation of the material. $F_{\mathbf{q}s}$ is the mode-specific deviation function, representing the difference of non-equilibrium phonon population of mode $\mathbf{q}s$, $n_{\mathbf{q}s}\left(= n_{\mathbf{q}s}^0 + n_{\mathbf{q}s}^0\left(n_{\mathbf{q}s}^0 + 1\right)\frac{dT}{dx}F_{\mathbf{q}s}\right)$, from equilibrium population $n_{\mathbf{q}s}^0$. The deviation function can be solved from the PBTE, which describes the balance between phonon diffusion and phonon scatterings due to various scattering mechanisms. Here, we consider the three-phonon scattering due to the anharmonicity of interatomic forces, phonon-boundary scattering, phonon-alloy scattering and phonon-nanodomain scattering, and the corresponding PBTE is expressed as[30,32]

$$v_{\mathbf{q}s}^\alpha \frac{\partial n_{\mathbf{q}s}^0}{\partial T} = \sum_{\mathbf{q}'s',\mathbf{q}''s''}\left[W_{\mathbf{q}s,\mathbf{q}'s'}^{\mathbf{q}''s''}\left(F_{\mathbf{q}''s''}^\alpha - F_{\mathbf{q}'s'}^\alpha - F_{\mathbf{q}s}^\alpha\right) + \frac{1}{2}W_{\mathbf{q}s}^{\mathbf{q}'s',\mathbf{q}''s''}\left(F_{\mathbf{q}''s''}^\alpha + F_{\mathbf{q}'s'}^\alpha - F_{\mathbf{q}s}^\alpha\right)\right] + \sum_{\mathbf{q}'s'}W_{\mathbf{q}s,\mathbf{q}'s'}^{\text{Alloy}}\left(F_{\mathbf{q}'s'}^\alpha - F_{\mathbf{q}s}^\alpha\right) + \sum_{\mathbf{q}'s'}W_{\mathbf{q}s,\mathbf{q}'s'}^{\text{ND}}\left(F_{\mathbf{q}'s'}^\alpha - F_{\mathbf{q}s}^\alpha\right) - \frac{n_{\mathbf{q}s}^0\left(n_{\mathbf{q}s}^0+1\right)F_{\mathbf{q}s}^\alpha}{L/2\left|v_{\mathbf{q}s}^\alpha\right|}, \quad (2)$$

where $W_{\mathbf{q}s,\mathbf{q}'s'}^{\mathbf{q}''s''}$ and $W_{\mathbf{q}s}^{\mathbf{q}'s',\mathbf{q}''s''}$ are the transition probabilities for three-phonon annihilation and decay processes, and $W_{\mathbf{q}s,\mathbf{q}'s'}^{\text{Alloy}}$ and $W_{\mathbf{q}s,\mathbf{q}'s'}^{\text{ND}}$ are the transition probabilities for phonon-alloy and phonon-nanodomain scattering processes. The last term in Eq. (2) represents the phonon-boundary scattering due to the limited length of the sample $L$.

Since atoms in alloy are randomly distributed, it is ambiguous to define the phonon dispersion and other phonon properties of the alloy that are required in Eq. (1) and Eq. (2). The virtual crystal approximation is employed to take into account the alloy effect,[33] where the Mo or W atoms in the alloy are replaced by virtual atoms with an atomic mass of $(1 - x)M_{\text{Mo}} + xM_{\text{W}}$. Here the lattice parameter and inter-atomic force constants of the virtual crystal are set according to the composition average of the quantities possessed by $MoS_2$ and $WS_2$. The optimized crystal structures and the inter-atomic force constants of both $MoS_2$ and $WS_2$ are obtained from the first-principles calculations. The phonon dispersion of the virtual crystal is then computed with the obtained atomic masses and the second-order harmonic force constants of the virtual crystal. The numerical details of the first-principles calculations and the calculated phonon dispersion of $Mo_{1-x}W_xS_2$ alloy are presented in Supplemental Material S1.[34]

Under the virtual crystal approximation, the transition probabilities, $W_{\mathbf{q}s,\mathbf{q}'s'}^{\mathbf{q}''s''}$ and $W_{\mathbf{q}s}^{\mathbf{q}'s',\mathbf{q}''s''}$, in Eq. (2) can be straightforwardly computed using the Fermi's golden rule with the third-order force constants of the virtual crystal as inputs. The expressions of $W_{\mathbf{q}s,\mathbf{q}'s'}^{\mathbf{q}''s''}$ and $W_{\mathbf{q}s}^{\mathbf{q}'s',\mathbf{q}''s''}$ are given in Ref. [30].

## B. Green's function approach for phonon-alloy and phonon-nanodomain scattering rates

Unlike the calculation of inelastic phonon-phonon scattering rates which is quite standard practice these days, it is more challenging to estimate the strengths of elastic scatterings, including phonon-alloy scatterings and phonon-nanodomain scatterings, since they are very sensitive to the detailed atomic configurations of the lattice imperfections. Compared to the virtual crystal, the randomly distributed Mo and W atoms in the alloy serve as independent impurities of the virtual crystal, which scatter phonons. We note that the distribution of the impurities might lead to phonon interference effects which can result in a different thermal conductivity, but such effects are not considered here. In this work, we employed an assumption of random and uniform distribution of impurities, similar to the many previous theoretical works for the prediction of the thermal conductivity of a few alloys[35,36] where the theoretical prediction were shown to be consistent with experimental measurements. The scattering strength due to phonon-alloy scattering, $W_{\mathbf{q}s,\mathbf{q}'s'}^{\text{Alloy}}$, can then be decomposed into two parts, one due to the Mo impurities, $W_{\mathbf{q}s,\mathbf{q}'s'}^{\text{Alloy,Mo}}$, and the other due to the W impurities, $W_{\mathbf{q}s,\mathbf{q}'s'}^{\text{Alloy,W}}$. When $WS_2$ nanodomains are embedded in the alloy, the W clusters in the nanodomains become another type of scatter centers in the virtual crystal leading to the scattering with the rate of $W_{\mathbf{q}s,\mathbf{q}'s'}^{\text{ND}}$. The main effects of these three types of scattering centers result in the mass and force field variation in the virtual crystal, both of which serve as the perturbations on the lattice vibration of the virtual crystal. Since the difference of the bonding stiffness in $MoS_2$ and $WS_2$ is quite small (~4%)[13] compared with the mass difference between Mo and W atoms (~90%), we expect that the phonon scattering rate caused by the effects of force field variance is much smaller compared with that caused by mass variance. In fact, we can make a simple estimation on the ratio between phonon scattering rates due to the difference in the bonding stiffness and due to the atomic mass difference. According to the Klemens' theory[37] for a point defect with respect to atom $i$, the ratio is expressed as $2[((g_i - g)/g)/((M_i - M)/M)]^2$, where $g_i$ is the average stiffness constant of the neareast-neighbor bonds from the impurities to the host lattice, $g$ is the average stiffness constant for host atoms, $M_i$ and $M$ are the mass of the impurity and the host atom. From the simple calculation, the ratio is found to be smaller than 0.5% when a W atom is inserted into $MoS_2$. To consider the force field variation after nanodomains are included, one has to perform the first-principles calculations with large supercells or employ higher-order force constant model[38] to obtain the interatomic force constants with respect to the atoms in/near the nanodomains. Considering the weak scattering caused by force field variance, in this work only the mass-difference-induced phonon scattering is taken into account when evaluating the phonon scattering rates due to alloy disorder and nanodomains.

We employ the Green's function approach to calculate the phonon scattering rates due to alloying and nanodomain embedding. This method takes into account fully the changes of dynamical matrix of the medium (virtual crystal) when scattering centers are introduced and thus could accurately estimate the strengths of elastic scatterings, including the phonon-alloy and phonon-nanodomain scatterings, as those applied to study the elastic phonon scattering due to Si/Ge nanoparticles in SiGe alloy[39] and due to vacancy defects in diamond.[40] The phonon scattering strength due to a specific type of scattering centers, $j$, is given by[39]

$$W^j_{\mathbf{q}s,\mathbf{q}'s'} = N^j \frac{\Omega\pi}{2\omega^2_{\mathbf{q}s}} \left| \langle \mathbf{q}s | \mathbf{T}_j(\omega^2_{\mathbf{q}s}) | \mathbf{q}'s' \rangle \right|^2 \delta(\omega_{\mathbf{q}s} - \omega_{\mathbf{q}'s'}), \tag{3}$$

where $N^j (= f^j / A^j)$ is number density of $j$-type scattering centers with the area fraction of the scattering centers $f^j$ and the total area of the unit cells occupied by each scatter center $A_j$, $\Omega$ is the volume into which the phonon eigenstates $|\mathbf{q}s\rangle$ are normalized, and $\mathbf{T}_j$ is the scattering matrix corresponding to the $j$-type scattering centers. While the area fraction for nanodomains is $f^{\mathrm{ND}}$, the area fraction is $(1-x)(1-f^{\mathrm{ND}})$ and $x(1-f^{\mathrm{ND}})$ for Mo impurities and W impurities, respectively. The scattering matrix, $\mathbf{T}$, is related to the perturbation matrix, $\mathbf{V}$, which is the difference between the dynamical matrix of the alloy embedded with nanodomains (perturbated crystal) and that of the virtual crystal one (unperturbated crystal), and the Green's function of the virtual crystal, $\mathbf{g}^+$. The $T$ matrix is expressed as

$$\mathbf{T}_j(\omega^2) = \left[ \mathbf{I} - \mathbf{V}_j \mathbf{g}^+ \right]^{-1} \mathbf{V}_j. \tag{4}$$

Since only the mass difference is considered and the force field difference is ignored in this work, all of the non-zero elements of the perturbation matrix are the diagonal elements corresponding to the $i$-th degrees of freedom associated with the atom $\tau$ that has a different mass $M_\tau$ from the virtual atom with a mass of $M_0$. The non-zero elements of the matrix is given by[39]

$$\mathbf{V}_{ii} = -\frac{M_\tau - M_0}{M_0} \omega^2. \tag{5}$$

The Green's function component with respect to any two degrees of freedom $i$ and $j$, is written as[28]

$$\mathbf{g}^+_{ij}(\omega^2) = \lim_{z \to \omega^2 + i0^+} \sum_{\mathbf{q}s} \frac{\langle i | \mathbf{q}s \rangle \langle \mathbf{q}s | j \rangle}{z - \omega^2_{\mathbf{q}s}}. \tag{6}$$

The Green's function of unperturbed 3-D crystals can be numerically evaluated by the analytical tetrahedron method proposed by Lambin and Vigneron.[41] In this work we extend their method for 2-D crystals as detailed in Supplemental Material S2.[34]

With all scattering terms in Eq. (2) are determined, the set of linear equations Eq. (2), with respect to $F_{qs}^{\alpha}$, is then self-consistently solved using the iterative method. Here we employed the biconjugate gradient stabilized method,[42] a variant of the conjugate gradient algorithm, to iteratively solve Eq. (2). In our calculation, we use $65\times65\times1$ q-points to sample the reciprocal space for phonon scattering and thermal conductivity calculations respectively, which ensures the presented in-plane thermal conductivity data are converged, with less than 2% difference if the meshes are further refined.

### III. Results and Discussion

### A. Low thermal conductivity of $Mo_{1-x}W_xS_2$ alloy embedded with nanodomains

With the phonon scattering rates due to alloying and nanodomain embedding calculated, the thermal conductivity of the alloy embedded with nanodomains can then be straightforwardly calculated using PBTE, i.e., Eq. (2). Figure 2(a) shows the calculated thermal conductivity of $Mo_{1-x}W_xS_2$ alloy with a sample length of 10 μm. It is found that the thermal conductivity reduces from 155 W/mK to around 25 W/mK when the composition of W, $x$, increases from 0% to 30% while the thermal conductivity is almost unchanged as $x$ is in the range between 30% and 70%. Such a U-shape dependence of thermal conductivity on the composition is quite typical for alloys, as seen in 3D alloys such as $Si_{1-x}Ge_x$[33] and $Mg_2Si_{1-x}Sn_x$.[43]

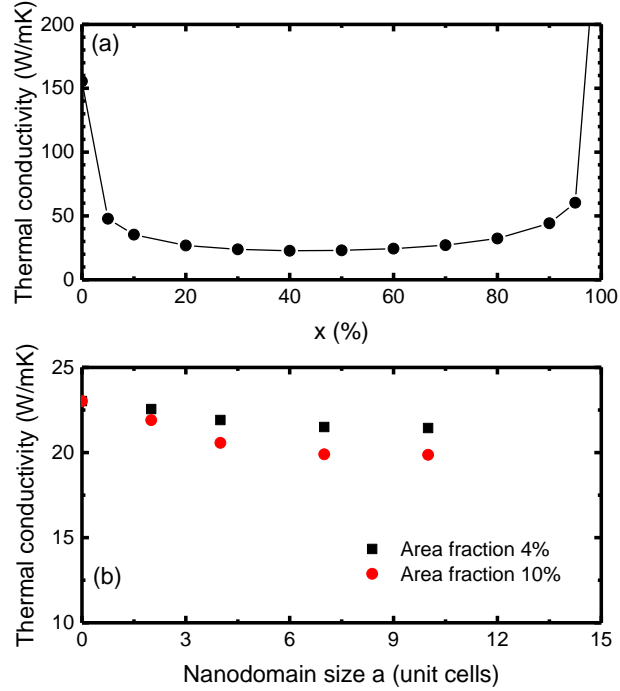

*Figure 2. (a) Thermal conductivity of Mo$_{1-x}$W$_x$S$_2$ at 300K as a function of alloy composition x. (b) Thermal conductivity of Mo$_{1-x}$W$_x$S$_2$ alloy embedded with W nanodomains as a function of the size of nanodomains.*

By embedding triangular WS$_2$ nanodomains into alloy, which serve as additional scattering centers for low-frequency phonons, the thermal conductivity of the alloys is further reduced. Figure 2(b) shows the dependence of thermal conductivity of Mo$_{0.5}$W$_{0.5}$S$_2$ alloy embedded with WS$_2$ nanodomains as a function of the area fraction and the size of nanodomains. The obtained thermal conductivity plotted as solid symbols, and is found to decrease with the size and area fraction of nanodomains when the area fraction increases from 0 to 10% and the size of nanodomains is changed from 1 to 10 unit cells. Apparently, embedding nanodomains considerably reduces the thermal conductivity below the alloy limit. For example, an additional 16 % reduction is observed when the area fraction of nanodomains $f^{ND}$ is 10% and the side length of each nanodomain *a* is 10 unit cells. Although the degree of thermal conductivity reduction of Mo$_{1-x}$W$_x$S$_2$ alloy due to nanodomains is smaller than some 3-D alloys, where additional factor of 2 to 4 is found when nanoparticles are embedded,[26] but it is comparable to nanostructured BiSbTe alloy.[27] In fact, the nanodomain/nanoparticle induced thermal conductivity reduction could be affected by many factors, including the alloy composition, crystal structures and so on. The dimensionality might be another factor that affects the thermal conductivity reduction, which calls for further investigation.

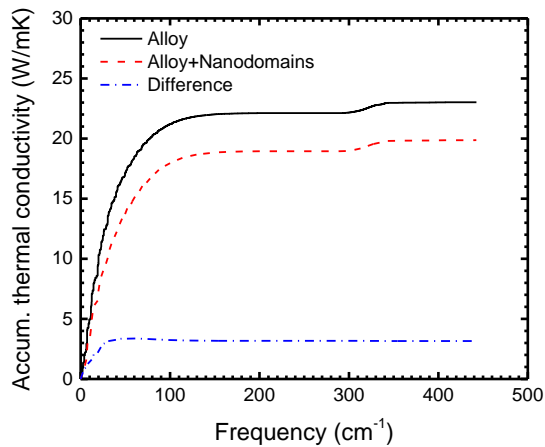

*Figure 3. The accumulative thermal conductivity of Mo$_{0.5}$W$_{0.5}$S$_2$ with and without nanodomains embedded as a function phonon frequency. The thermal conductivity difference is mainly caused by low-frequency phonons.*

To gain insights about the thermal conductivity reduction when nanodomains are embedded in an alloy matrix, we examine the accumulated thermal conductivity of Mo$_{0.5}$W$_{0.5}$S$_2$ alloy and the alloy embedded with nanodomains whose area fraction and side length are 10% and 10 unit cells as a function of phonon frequency, as shown in Fig. 3. It is clearly seen that the thermal conductivity difference between the alloy with and without nanodomains is mainly caused by low-frequency phonons. This indicates that the long-wavelength phonons are more effectively scattered by the nanodomains than by the alloy components that are replaced by the nanodomains.

## B. Phonon-nanodomain scattering rates

To understand the scattering mechanisms of long-wavelength/low-frequency phonons in the alloy embedded with nanodomains, we present the calculated phonon-nanodomain scattering rates of longitudinal acoustic phonons of single-layer $Mo_{1-x}W_xS_2$ on a high symmetry line from the $\Gamma$ point, $(0, 0)$, to the M point, $(0, \pi/a_0\sqrt{3})$, as a function of phonon frequency in the log-log scale, which is shown in Fig. 4(a). The scattering rate of a specific mode due to the nanodomains is written as

$$\Gamma_{\mathbf{q}s} = \sum_{\mathbf{q}'s'} W^{\mathrm{ND}}_{\mathbf{q}s,\mathbf{q}'s'} / n^0_{\mathbf{q}s}(n^0_{\mathbf{q}s} + 1), \qquad (6).$$

Here the scattering rates are normalized by the area fraction of nanodomains in order to fairly compare the effects of nanodomains with different sizes on phonon scattering. As expected, the strength of phonon-nanodomain scattering is greatly affected by both the phonon frequency and the size of nanodomains. For low-frequency phonons, whose wavelengths are much larger than the characteristic size of nanodomains, the frequency-dependent scattering rate follows a nice $\omega^{-3}$ scaling relation, which is consistent with the characteristics of Rayleigh scattering in 2-D crystals.[29] The scattering rate also increases with the size of nanodomains for these long-wavelength/low-frequency phonons. Figure 4(b) shows phonon-nanodomain scattering rates of long-wavelength acoustic phonons with a wavevector of $0.03(0, \pi/a_0\sqrt{3})$ versus the number of W atoms in each nanodomain, $N$, where the scattering rate is found to be proportional to $N$.

In fact, the phonon scattering rate of the Rayleigh scattering can be easily obtained using the Born approximation, where the **T** matrix in Eq. (3) is replaced by the perturbation matrix **V.** The obtained total scattering rate of a specific mode due to the nanodomains scales as

$$\Gamma_{\mathbf{q}s} \propto \frac{1}{N}\left(N\frac{\Delta M}{M}\right)^2. \tag{7}$$

From this simple relation between the scattering rate, $\Gamma$, and the number of W atoms in each nanodomain, $N$, one can expect that the ratio between the scattering rate due to nanodomains and that due to Mo and W impurities of alloy, which can be regarded as nanodomains with 1 atom, should scale proportionally with $N$. As a result, the long-wavelength phonons could be more strongly scattered in the alloy embedded with nanodomains than the alloy only.

On the other hand, Fig. 4(a) shows that the scattering rate of high-frequency phonons, whose wavevectors are close to the edge of the first Brillouin zone, decreases with the size of nanodomain, which is drastically different from the low-frequency phonons. Figure 4(b) shows the phonon scattering rates for the acoustic phonons at (0, 0.97M), which are taken as examples of short-wavelength phonons. The phonon scattering rate clearly follow $\Gamma \propto 1/\sqrt{N}$. This is because it is at the geometric scattering regime[44,45] where the scattering cross section is limited by the size of the nanodomain when the wavelength is smaller than the size of nanodomain.[44,46] Since the scattering rate of the mode $\mathbf{q}s$ is linked to the scattering cross section through $\Gamma_{\mathbf{q}s} = N^{\text{ND}}\sigma_{\mathbf{q}s}v_{\mathbf{q}s}$ in the 2-D systems, where $\sigma$ is the scattering cross section, $v$ is the phonon group velocity and $N^{\text{ND}}$ is the number density of the scatterers, the scattering rate is inversely proportional to the scattering cross section, or equivalently the side length, $a$, of nanodomains, which scales as $\sqrt{N}$.

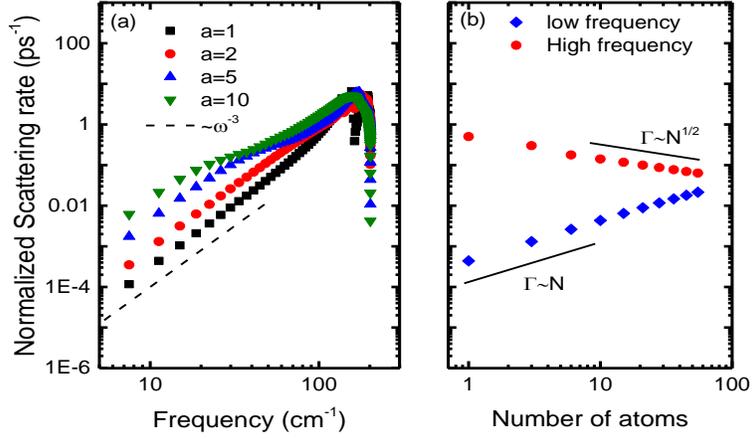

*Figure 4. (a) The phonon-nanodomain scattering rates of LA phonons from the Γ point to the M point as a function of phonon frequency. (b) The phonon-nanodomain scattering rates of long-wavelength LA phonon at 0.03(0, π/a₀√3) and short-wavelength LA phonon at 0.97(0, π/a₀√3). The scattering rates are normalized by the area fraction of the nanodomains.*

Because the strengths of phonon scattering for long-wavelength and short-wavelength phonons follow opposite trends with the size of nanodomains, there exists a minimum thermal conductivity with optimal size of nanodomains. Due to the computational limitations, we are not able to directly evaluate the thermal conductivity of the samples embedded with nanodomains using the Green's function approach when the size of nanodomains is larger than 10 unit cells. However, as seen in Fig. 2(b), the optimal size of nanodomains could be around 10 unit cells, where the thermal conductivity of alloy embedded with nanodomains is almost unchanged with the size.

## C. Thermal conductivity dependence of alloy composition

Figure 5 shows the dependence of thermal conductivity of alloy embedded with nanodomains whose size is 10 unit cells and area fraction is 10% on the composition of alloy matrix. The thermal conductivity value changes non-monotonically with the composition of alloy matrix, just as the alloy without embedding nanodomains. But the minimum thermal conductivity occurs at the composition of $x \approx 20\%$, while that of pure alloy is at $x \approx 50\%$. This means that the pure alloy with the minimum thermal conductivity is not necessarily the best matrix for such nanocomposites to achieve lowest thermal conductivity. For example, the thermal conductivity of alloy embedded with 10-unit-cell-large $WS_2$ nanodomains at 10% area fraction decreases from 19.8 W/mK to 16.9 W/mK when the composition of W changes from 50% to 20%.

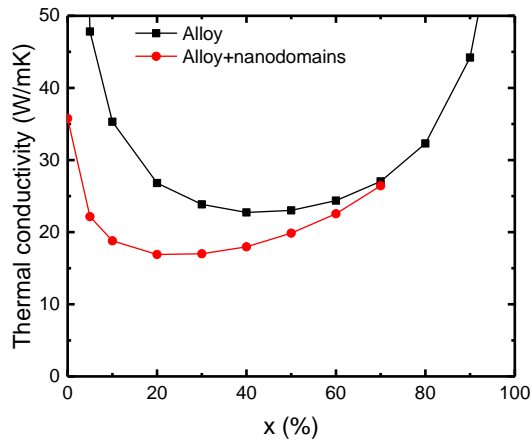

*Figure 5. Thermal conductivity of $Mo_{1-x}W_xS_2$ alloy embedded with 10-unit-cell-large $WS_2$ nanodomains at 10% area fraction at 300 K as a function alloy composition x.*

To understand the lower thermal conductivity in $Mo_{0.8}W_{0.2}S_2$-based nanocomposites than that in $Mo_{0.5}W_{0.5}S_2$-based nanocomposites with the same $WS_2$ nanodomains, we compare the phonon-alloy scattering rates and the phonon-nanodomain scattering rates in different alloy compositions, as shown in Fig. 6(a) and (b), respectively. As expected, it is seen from Fig. 6(a) that the phonon-alloy scattering rates in $Mo_{0.8}W_{0.2}S_2$-based nanocomposite become smaller than those in $Mo_{0.5}W_{0.5}S_2$-based one due to less disorder in $Mo_{0.8}W_{0.2}S_2$ alloy-based nanocomposites. However,

Fig. 6(b) shows that the scattering rates due to nanodomains become almost doubled in $Mo_{0.8}W_{0.2}S_2$ alloy-based nanocomposites for low-frequency phonons, but the scattering rates are almost unchanged for high-frequency phonons. This is expected because the scattering rates for the low-frequency phonons are mainly determined by the total mass contrast, $(N\Delta M/M)$, between the nanodomain and the averaged mass of alloy matrix, as indicated by the Born approximation, Eq. (6), while for high-frequency phonons, the scattering rates are governed by the geometrical size of nanodomains, which is almost unchanged with respect to the alloy composition. Since most of heat is conducted by low-frequency phonons in the alloy and nanocomposites, as indicated from Fig. 4, the enhanced scattering for low-frequency phonons due to nanodomains can effectively reduce the total conductivity of alloys embedded with nanodomain.

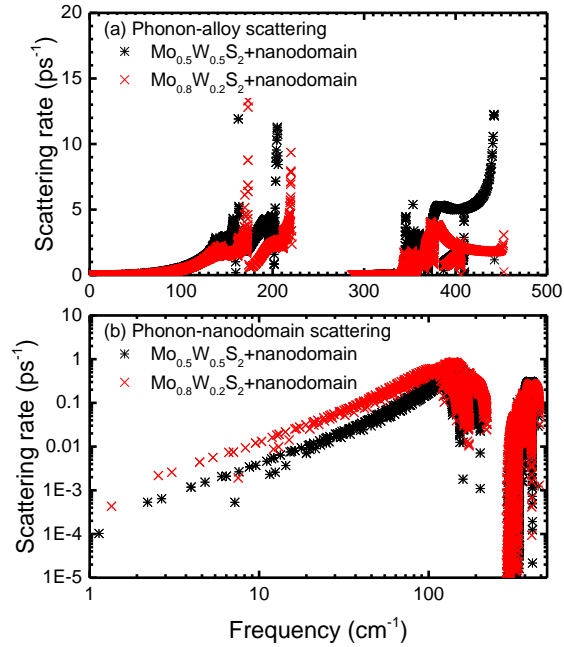

*Figure 6. (a) Phonon-alloy scattering rates as a function of phonon frequency. (b) Phonon-nanodomain scattering rates as a function of phonon frequency.*

### D. Strategies to further reduce the thermal conductivity

In this work, we have demonstrated a possible way to significantly reduce the thermal conductivity of single-layer $MoS_2$ through alloying and nanodomain embedding.

There are indeed plenty of rooms to further tune the thermal conductivity. In current study, all nanodomains are of the same size, but it is expected that embedding nanodomains with different sizes could help to reduce further the thermal conductivity. Another possibility is to either introduce $MoSe_2$ and/or $WSe_2$ components into the $Mo_{1-x}W_xS_2$ alloy or use them as nanodomains. Because the mass ratio between Se and S atoms is larger than that between W and Mo atoms and the force field difference between the $MoS_2$ and $MoSe_2$ is also larger than that between $MoS_2$ and $WS_2$,[13] alloying and introducing nanodomains on chalcogen atoms can enhance the phonon scattering in the 2-D alloy compared with the case that only the metal atoms are mixed. Once the monolayers after nanostructuring are obtained, one might be able to stack them together to form a layered material with very low thermal conductivity. Recent study showed that by intercalation layered materials with organic components, the thermoelectric properties can be significantly improved for layered materials mainly due to the lower thermal conductivity.[6] It might be possible to further optimize the obtained layered material through intercalation. Despite these possible approaches to further reduce the thermal conductivity of 2-D materials, it should be noticed that the parameters that determine the thermoelectric performance of the material are usually highly correlated. In order to optimize the thermoelectric performance of the nanostructured 2-D materials, more detailed analysis should be conducted, especially on the electronic properties.

## IV. Summary

In summary, we study the phonon scattering mechanisms and the thermal conductivity in $Mo_{1-x}W_xS_2$ alloy embedded with triangular $WS_2$ nanodomains using first-principles-based PBTE calculations. The phonon scattering rates due to alloying and nanodomain embedding are evaluated using the Green's function approach. The thermal conductivity of $Mo_{1-x}W_xS_2$ is found to be significant lower than that of pristine $MoS_2$. For example, the thermal conductivity of $Mo_{0.5}W_{0.5}S_2$ is only 16% of $MoS_2$ when the sample length is 10 μm. By embedding $WS_2$ nanodomains into $Mo_{0.8}W_{0.2}S_2$, the obtained thermal conductivity of alloy can be only about 1/10 of that of $MoS_2$ due to both the strong alloy scattering for high-frequency phonons and nanodomain scattering for low-frequency phonons. If the electronic properties of these nanocomposites can be retained to be similar as $MoS_2$, $Mo_{1-x}W_xS_2$ alloy embedded with $WS_2$ could be a good thermoelectric material, which is worth of further investigation.

## Acknowledgements

This work was supported by the National Science Foundation (Award No. 1511195). XG acknowledges the Teets Family Endowed Doctoral Fellowship. This work utilized the Janus supercomputer, which is supported by the National Science Foundation (Award No. CNS-0821794), the University of Colorado Boulder, the University of Colorado Denver, and the National Center for Atmospheric Research. The Janus supercomputer is operated by the University of Colorado Boulder.